\documentclass{article}

\usepackage[nonatbib, preprint]{neurips_2023}

\usepackage[utf8]{inputenc} % allow utf-8 input
\usepackage[T1]{fontenc}    % use 8-bit T1 fonts
\usepackage{hyperref}       % hyperlinks
\usepackage{url}            % simple URL typesetting
\usepackage{booktabs}       % professional-quality tables
\usepackage{amsfonts}       % blackboard math symbols
\usepackage{nicefrac}       % compact symbols for 1/2, etc.
\usepackage{microtype}      % microtypography
\usepackage{xcolor}         % colors

\usepackage{amsmath, bm}
\usepackage{amssymb}
\usepackage{mathtools}
\usepackage{amsthm}
\usepackage{xspace}
\usepackage{multirow}
\usepackage{pifont}

\newcommand{\name}{\texttt{Uni-Mol+}\xspace}
\newcommand{\sa}{\texttt{SelfAttentionPairBias}\xspace}
\newcommand{\opm}{\texttt{OuterProduct}\xspace}
\newcommand{\tm}{\texttt{TriangularUpdate}\xspace}
\newcommand{\ffn}{\texttt{FFN}\xspace}
\newcommand{\embedding}{\texttt{Embedding}\xspace}

\title{Highly Accurate Quantum Chemical Property Prediction with Uni-Mol+}

\author{Shuqi Lu$^1$, Zhifeng Gao$^1$, Di He$^2$, Linfeng Zhang$^1$, Guolin Ke$^1$ \\
$^1$DP Technology\\
$^2$Peking University\\
\{lusq, gaozf\}@dp.tech \\ 
 dihe@pku.edu.cn, \{zhanglf, kegl\}@dp.tech \\
}

\begin{document}

\maketitle

\begin{abstract}
Recent developments in deep learning have made remarkable progress in speeding up the prediction of quantum chemical (QC) properties by removing the need for expensive electronic structure calculations like density functional theory. 
However, previous methods learned from 1D SMILES sequences or 2D molecular graphs failed to achieve high accuracy as QC properties primarily depend on the 3D equilibrium conformations optimized by electronic structure methods, far different from the sequence-type and graph-type data. 
In this paper, we propose a novel approach called \name to tackle this challenge. \name first generates a raw 3D molecule conformation from inexpensive methods such as RDKit. Then, the raw conformation is iteratively updated to its target DFT equilibrium conformation using neural networks, and the learned conformation will be used to predict the QC properties. 
To effectively learn this update process towards the equilibrium conformation, we introduce a two-track Transformer model backbone and train it with the QC property prediction task. We also design a novel approach to guide the model's training process. Our extensive benchmarking results demonstrate that the proposed \name significantly improves the accuracy of QC property prediction in various datasets. 
We have made the code and model publicly available at \url{https://github.com/dptech-corp/Uni-Mol}.
\end{abstract}

\section{Introduction}

The design of novel materials and molecules through computational methods heavily relies on predicting quantum chemical (QC) properties, such as free energy and the HOMO-LUMO gap. High-throughput screening on a large database is the most straightforward approach to discovering molecules with desired properties~\cite{jain2013commentary}. However, this approach can be cost-prohibitive due to using electronic structure methods, such as density functional theory (DFT)~\cite{jones2015density}, which can take several hours to calculate properties on a single molecule. Therefore, finding ways to reduce the costs of calculating QC properties is crucial. 

Recent studies have demonstrated the potential of using deep learning to accelerate QC property prediction~\cite{luo2022one, ying2021transformers, shi2022benchmarking}. This involves training a deep neural network model to predict the property using molecular inputs, thus replacing the heavy DFT calculations. Previous research has mainly utilized 1D SMILES \cite{wang2019smiles, ross2022large, gomez2018automatic} sequences or 2D molecular graphs \cite{gilmer2017neural, luo2022your, kim2022pure, park2022grpe, hussain2022global, ying2021transformers} as molecular inputs due to their easy obtainability. However, predicting QC properties from 1D SMILES and 2D molecular graphs can be ineffective since most QC properties are estimated from the 3D equilibrium conformations optimized by DFT, which has an inherent and significant discrepancy with the 1D/2D molecular format. 

To address this challenge, we propose a method called \name in this paper. Unlike previous approaches that directly predict QC properties from 1D/2D data, \name first generates raw 3D conformation from 1D/2D data using cheap methods, such as RDKit \cite{landrum2016rdkit}. As the raw conformation is inaccurate, \name then iteratively updates it towards the DFT equilibrium conformation using neural networks and predicts QC properties from the learned conformation. It is easy to see that the success of this learning paradigm highly depends on the quality of the learned conformation. To obtain accurate predictions, we use large-scale datasets (e.g.,  PCQM4MV2 benchmark) to build up millions of pairs of RDKit-generated raw conformation and high-quality DFT equilibrium conformation and learn the update process from this supervised information. With a carefully designed model backbone and training strategy, \name shows superior performance in various benchmarks.

\vspace{-6pt}
\paragraph{Model Backbone} 
We adopted a two-track Transformer backbone, similar to Uni-Mol \cite{zhou2022uni}. The backbone consists of an atom representation track and a pair representation track.
Besides, we propose several modifications to enhance the model capacity further. These include:
i) An enhanced pair representation that is initialized with 3D spatial positional encoding and graph positional encoding. Subsequently, it is repeatedly updated via three operators: an outer product of atom representation, a triangular update operator, and a feed-forward network.
ii) An iterative process is employed to continuously update the 3D coordinates, towards the DFT equilibrium conformation.

\vspace{-6pt}
\paragraph{Training Strategy}
L1 loss is used to train the QC properties. For the learning of the conformation update process, we introduce a novel training strategy. We sample conformations from the trajectory between the RDKit-generated raw conformation and the DFT equilibrium conformation, and use the sampled conformation as input to predict the equilibrium conformation. It is crucial to note that the actual trajectory is often unknown in many datasets; therefore, we utilize a pseudo trajectory that presumes a linear process between two conformations. Furthermore, we devise a sampling strategy for obtaining conformations from the pseudo trajectory to serve as the model's input during training. 
This strategy addresses the distributional shift between training and inference and enhances the learning of an accurate mapping from the equilibrium comformation to the QC properties. Details can be found in Section \ref{sec:training}.

\vspace{-6pt}
\paragraph{Experimental Result}

Our proposed \name is evaluated on several public benchmark datasets, including PCQM4MV2~\cite{DBLP:conf/nips/HuFRNDL21} and Open Catalyst 2020 (OC20)~\cite{chanussot2021open}. In all these benchmarks, \name outperforms all previous state-of-the-art methods by a significant margin. Furthermore, we conducted an ablation study to examine the necessity of the components in \name.  
The results provide clear evidence of the effectiveness of the proposed \name.

 \begin{figure*}[!tbp]
 %\vspace{-4pt}
 \centering
\includegraphics[width=0.9\linewidth]{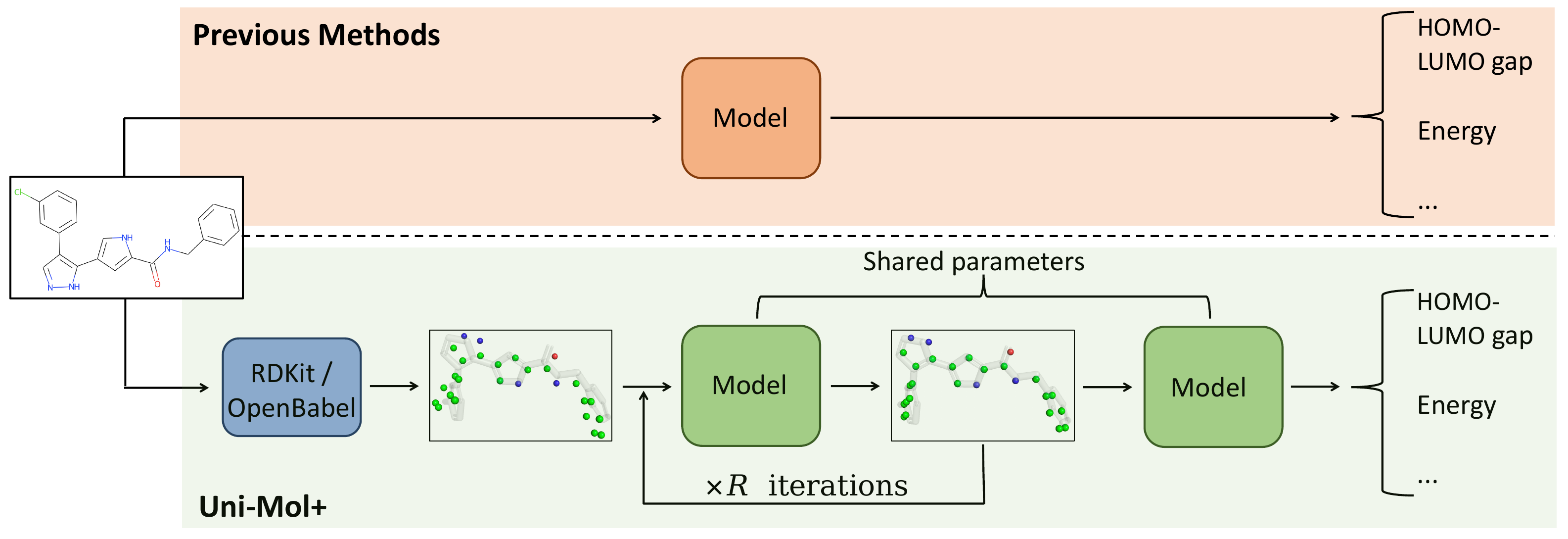}
\vspace{-10pt}
\caption{In contrast to prior methods that directly predict QC properties from 1D/2D data, \name uses a different approach. It first generates raw 3D conformation from 1D/2D data using cheap tools like RDKit, and then iteratively updates it towards the DFT equilibrium conformation. Finally, it predicts QC properties using the learned conformation. } \label{fig:pcq_model_compare}
\end{figure*}

\section{Related Work}
\vspace{-6pt}
\paragraph{Deep QC property prediction models with 1D/2D information}
Some prior works have solely utilized 1D information, such as SMILES sequences, to make predictions. These include Wang et al.~\cite{wang2019smiles}, Ross et al.~\cite{ross2022large}, and Gomez et al.~\cite{gomez2018automatic}. However, to incorporate more information, a number of studies have employed 2D graphs or fingerprints derived from SMILES sequences. Gilmer et al.~\cite{gilmer2017neural} utilized Graph Neural Network (GNN) to learn the molecular graph representation, while Graphormer~\cite{ying2021transformers} extended Transformer models to graph tasks through graph structural encodings. Numerous subsequent studies~\cite{gilmer2017neural, luo2022your, kim2022pure, park2022grpe, hussain2022global} have employed Transformer models on graph tasks and have made significant improvements and innovations by including self-attention mechanisms and relative positional encoding, thereby enhancing the capabilities of Transformer models in graph tasks. 

\vspace{-6pt}
\paragraph{Deep QC property prediction models with 3D information}
As 3D information is critical in predicting QC property, several recent studies have incorporated 3D information.
Some works use 3D structure information to enhance the 2D representation during training. St{\"a}rk et al.~\cite{stark20223d} maximize the mutual information between 3D vectors and the 2D representations of a GNN to make them contain latent 3D information, and Liu et al.~\cite{liu2021pre} leverage the correspondence and consistency between 2D topological structures and 3D geometric views to learn a 2D molecular graph encoder enhanced by richer and more discriminative 3D geometry. Zhu et al. propose a unified 2D and 3D pre-training scheme and make the 2D and 3D encoders promote each other. However, these works only implicitly embed 3D structural information into 2D representation, while only 2D information is used during inference. 
Luo et al.~\cite{luo2022one} propose Transformer-M, an encoder capable of handling 2D and 3D input formats. They are among the first to demonstrate that incorporating 3D information during training can enhance the performance of using 2D input alone in inference. However, the application of Transformer-M is restricted to taking 2D molecular graph or 3D DFT equilibrium conformation as input.
Some works take 3D conformation as input~\cite{schutt2017schnet, klicpera2003directional, hutchinson2021lietransformer, tholke2022equivariant, shi2022benchmarking, ying2021transformers, zhou2022uni} and consider the equivalence or invariance of rotation and translation in the model. 
For instance, Zhou et al.~\cite{zhou2022uni} introduce Uni-Mol, which utilizes the 3D conformation generated by RDKit as input and incorporates a 3D position recovery task for pretraining. Shi et al.~\cite{shi2022benchmarking} propose Graphormer-3D, an extension of Graphormer~\cite{ying2021transformers} to a 3D Transformer model. Graphormer-3D takes the initial 3D conformation provided by the OC20 dataset~\cite{chanussot2021open} and predicts the energy at equilibrium. While this setup is similar to \name, the model's backbone and the two models' training strategies differ significantly. 

\vspace{-6pt}
\paragraph{3D conformation optimization}
Optimizing molecular conformations towards an equilibrium state is a crucial challenge in computational chemistry. The most popular method for this task is Density Functional Theory (DFT), which offers high accuracy, but at a substantial computational cost. Several deep learning-based potential energy models, such as Deep Potential ~\cite{zhang2018deep}, have been proposed to address this issue. These models use neural networks to replace the costly potential calculations in DFT, thereby improving efficiency. However, deep potential models still require dozens or hundreds of steps to optimize the conformation iteratively based on the predicted potentials. In contrast, \name only requires a few rounds of optimization. Additionally, \name can end-to-end optimize the conformation, whereas deep potential models cannot.
Although other works \cite{guan2022energyinspired, zhou2022uni} also optimize RDKit-generated conformations towards DFT conformations, these works generally focus on benchmarking conformation rather than QC property. Furthermore, compared with \name, these works differ significantly in terms of model backbone and training strategy.

\section{Method}
For any molecule, \name first obtains a raw 3D conformation generated by cheap methods, such as template-based methods from RDKit and OpenBabel. It then learns the target conformation, i.e., the equilibrium conformation optimized by DFT, by an iterative update process from the raw conformation. In the final step, the QC properties are predicted based on the learned conformation. To achieve this goal, we introduce a new model backbone and a novel training strategy for updating conformation and predicting QC properties, discussed in subsequent subsections.

\subsection{Model Backbone}

\begin{figure*}[!tbp]
 %\vspace{-4pt}
 \centering
\includegraphics[width=0.95\linewidth]{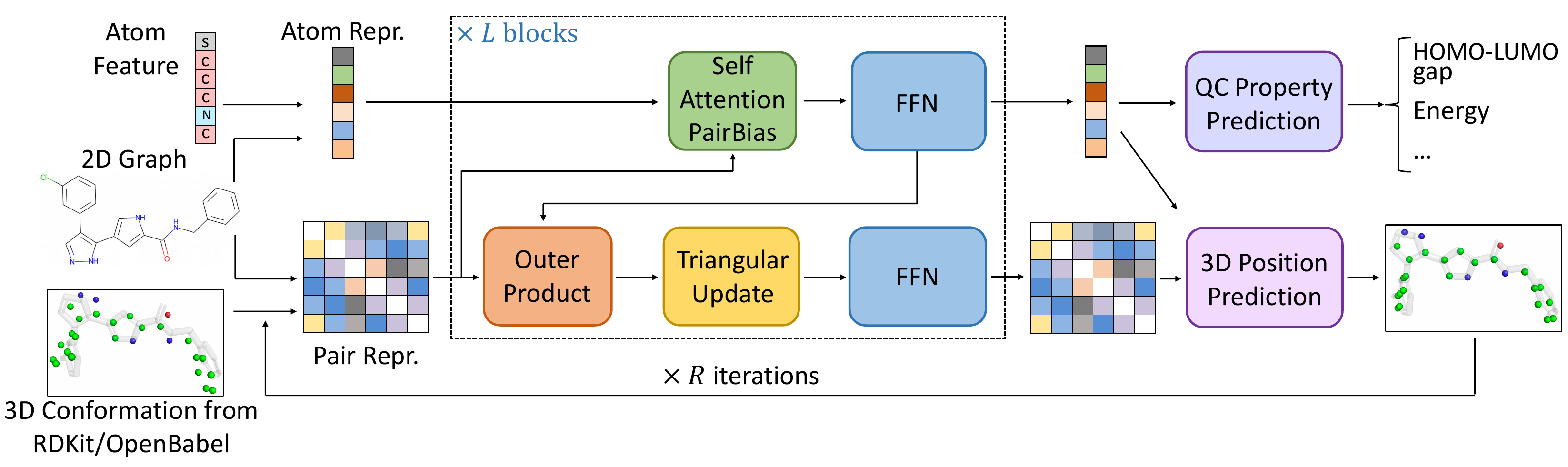}
\vspace{-10pt}
\caption{The \name  backbone consists of $L$ blocks, each of which maintains two tracks of representations - atom and pair, initialized by atom features and 2D graph/3D conformation, respectively. These representations communicate with each other at every block. Based on this backbone model, \name iteratively updates the raw conformation (i.e., 3D coordinates of atoms) towards the DFT equilibrium conformation for $R$ iterations.} \label{fig:model}
\end{figure*}

We design a novel model backbone that can learn the conformation and predict QC property simultaneously, denoted as $(y, \bm{\hat{r}}) = f(\bm{X}, \bm{E}, \bm{r} ; \bm{\theta})$.
This model uses atom features ($\bm{X} \in \mathbb{R}^{n \times d_f}$, where $n$ is the number of atoms and $d_f$ is atom feature dimension), edge features ($\bm{E} \in \mathbb{R}^{n \times n \times d_e}$, where $d_e$ is the edge feature  dimension), and 3D coordinates of atoms ($\bm{r} \in \mathbb{R}^{n \times 3}$), with learnable parameters $\bm{\theta}$, to predict a quantum property $y$ and updated 3D coordinates $\bm{\hat{r}} \in \mathbb{R}^{n \times 3}$.
Like the Uni-Mol model \cite{zhou2022uni}, two tracks of representations are maintained: 1) Atom representation ($\bm{x} \in \mathbb{R}^{n \times d_x}$, where $d_x$ is the dimension of the atom representation) and 2) Pair representation ($\bm{p} \in \mathbb{R}^{n \times n \times d_p}$, where $d_p$ is the dimension of the pair representation). The model has $L$ blocks, with $\bm{x}^{(l)}$ and $\bm{p}^{(l)}$ representing the $l$-th block's outputs.
The overall architecture of \name is depicted in Fig.~\ref{fig:model}. Compared with Uni-Mol, there are several key differences, which will be outlined in the following paragraphs.

\vspace{-6pt}
\paragraph{Positional Encoding}
Similar to previous works~\cite{ying2021transformers, zhou2022uni}, we use pair-wise encoding to encode the 3D spatial and 2D graph positional information. Specifically, for 3D spatial information, we utilize the Gaussian kernel for encoding, as done in previous studies~\cite{shi2022benchmarking, zhou2022uni}. The encoded 3D spatial positional encoding is denoted by $\bm{\psi}^{\text{3D}}$.

In addition to the 3D positional encodings, we also incorporate graph positional encodings similar to those used in Graphormer. This includes the shortest-path encoding, represented by $\bm{\psi}^{SP}_{i,j} = \embedding(\text{sp}_{ij})$ where $\text{sp}_{ij}$ is the shortest path between atoms $(i,j)$ in the molecular graph. Additionally, instead of the time-consuming multi-hop edge encoding method used in Graphormer, we utilize a more efficient one-hop bond encoding, denoted by $\bm{\psi}^{Bond} = \sum_{i=1}^{d_e}\embedding(\bm{E}_i)$, where $\bm{E}_{i}$ is the $i$-th edge feature. Combined above, the positional encoding is denoted as $\bm{\psi} = \bm{\psi}^{\text{3D}} + \bm{\psi}^{\text{SP}} + \bm{\psi}^{\text{Bond}}$. And the pair representation $\bm{p}$ is initialized by $\bm{\psi}$, i.e., $\bm{p}^{(0)} = \bm{\psi}$.

\vspace{-6pt}
\paragraph{Update of Atom Representation}
The atom representation $\bm{x}^{(0)}$ is initialized by the embeddings of atom features, the same as Graphormer. At $l$-th block, $\bm{x}^{(l)}$ is sequentially updated as follow:
\begin{equation}
\begin{aligned}
    \bm{x}^{(l)} = \bm{x}^{(l-1)} &+ \sa\left(\bm{x}^{(l-1)}, \bm{p}^{(l-1)} \right), \\
    \bm{x}^{(l)} = \bm{x}^{(l)} &+ \ffn\left(\bm{x}^{(l)} \right).
\end{aligned}
\end{equation}

The \sa function is denoted as:
\begin{equation}
\begin{aligned}
    \bm{Q}^{(l, h)} &= \bm{x}^{(l-1)} \bm{W}_Q^{(l, h)}; \quad \bm{K}^{(l, h)} = \bm{x}^{(l-1)} \bm{W}_K^{(l, h)}; \\
    {\color{blue} \bm{B}^{(l, h)}} &= \bm{p}^{(l-1)} \bm{W}_B^{(l,h)}; \quad \bm{V}^{(l, h)} = \bm{x}^{(l-1)} \bm{W}_V^{(l, h)}; \\
    \texttt{output} &=  \texttt{softmax} \left( \frac{ \bm{Q}^{(l,h)} (\bm{K}^{(l,h)})^T}{\sqrt{d_h}} + {\color{blue} \bm{B}^{(l, h)}} \right) \bm{V}^{(l, h)}; \\
\end{aligned}
\end{equation}

where $d_h$ is the head dimension, $\bm{W}_Q^{(l, h)}, \bm{W}_K^{(l, h)}, \bm{W}_V^{(l, h)} \in \mathbb{R}^{d_x \times d_h}$,  $\bm{W}_B^{(l, h)} \in  \mathbb{R}^{d_p \times 1}$.
\ffn is a feed-forward network with one hidden layer.  For simplicity, layer normalizations are omitted. Compared to the standard Transformer layer, the only difference here is the usage of attention bias term ${\color{blue}\bm{B}^{(l,h)}}$ to incorporate $\bm{p}^{(l-1)}$ from the pair representation track.

\vspace{-6pt}
\paragraph{Update of Pair Representation}
The pair representation $\bm{p}^{(0)}$ is initialized by the positional encoding $\bm{\psi}$. The update process of pair representation begins with an outer product of $\bm{x}^{(l)}$, followed by a $\mathcal{O}(n^3)$ triangular multiplication, and is then concluded with an \ffn layer. 
Formally, at $l$-th block, $\bm{p}^{(l)}$ is sequentially updated as follow:
\begin{equation}
\begin{aligned}
    \bm{p}^{(l)} &= \bm{p}^{(l-1)} + \opm(\bm{x}^{(l)}); \\
    \bm{p}^{(l)} &= \bm{p}^{(l)} + \tm(\bm{p}^{(l)}); \\
    \bm{p}^{(l)} &= \bm{p}^{(l)} + \ffn(\bm{p}^{(l)}).
\end{aligned}
\end{equation}
The \opm is used for atom-to-pair communication, denoted as :
\begin{equation}
\begin{aligned} 
    \bm{a} &= \bm{x}^{(l)} \bm{W}_{O1}^{(l)}, \bm{b}= \bm{x}^{(l)} \bm{W}_{O2}^{(l)}; \\
    \bm{o}_{i,j} &= \texttt{flatten} (\bm{a}_i \otimes \bm{b}_j); \\
    \texttt{output} &=  \bm{o} \bm{W}_{O3}^{(l)},  
\end{aligned}
\end{equation}
where $\bm{W}_{O1}^{(l)}, \bm{W}_{O2}^{(l)} \in \mathbb{R}^{d_x \times d_o}$, $d_o$ is the hidden dimension of \opm, and $\bm{W}_{O3}^{(l)} \in \mathbb{R}^{d_o^2 \times d_p}$, $\bm{o}=[\bm{o}_{i,j}]$. Please note that $\bm{a}, \bm{b}, \bm{o}$ are temporary variables in the \opm function. \tm is used to enhance pair representation further, denoted as:
\begin{equation}
\begin{aligned}
    \bm{a} &= \texttt{sigmoid} \left(\bm{p}^{(l)} \bm{W}_{T1}^{(l)}\right) \odot \left(\bm{p}^{(l)} \bm{W}_{T2}^{(l)}\right); \\
    \bm{b} &= \texttt{sigmoid} \left(\bm{p}^{(l)} \bm{W}_{T3}^{(l)}\right) \odot \left(\bm{p}^{(l)} \bm{W}_{T4}^{(l)}\right); \\
    \bm{o}_{i,j} &= \sum_{k} {\bm{a}_{i,k} \odot \bm{b}_{j,k}} + \sum_{k} {\bm{a}_{k,i} \odot \bm{b}_{k,j}};    \\
    \texttt{output} &=  \texttt{sigmoid} \left(\bm{p}^{(l)} \bm{W}_{T5}^{(l)}\right) \odot \left(\bm{o} \bm{W}_{T6}^{(l)} \right),
\end{aligned}
\end{equation}
where $\bm{W}_{T1}^{(l)}, \bm{W}_{T2}^{(l)}, \bm{W}_{T3}^{(l)}, \bm{W}_{T4}^{(l)} \in \mathbb{R}^{d_p \times d_t} $, $\bm{W}_{T5}^{(l)} \in \mathbb{R}^{d_p \times d_p}$, $\bm{W}_{T6}^{(l)} \in \mathbb{R}^{d_t \times d_p}$, $\bm{o}=[\bm{o}_{i,j}]$, and $d_t$ is the hidden dimension of \tm. $\bm{a}, \bm{b}, \bm{o}$ are temporary variables. The \tm is inspired by the Evoformer in AlphaFold \cite{jumper_highly_2021}. The difference is that AlphaFold uses two modules, ``outgoing'' ($\bm{o}_{i,j} = \sum_{k} {\bm{a}_{i,k} \odot \bm{b}_{j,k}}$) and ``incoming''  ($\bm{o}_{i,j} = \sum_{k} {\bm{a}_{k,i} \odot \bm{b}_{k,j}}$) respectively. In \name, we merge the two modules into one to save the computational cost. \looseness=-1

\vspace{-6pt}
\paragraph{Iterative Update of 3D Coordinates}

To update the 3D coordinates of atoms, we directly use the 3D prediction head proposed in Graphormer-3D \cite{shi2022benchmarking}.  The conformation optimization process in many practical applications, such as Molecular Dynamics, is iterative. This approach is also employed in the \name. The number of conformation update iterations denoted as $R$, is a hyperparameter. 
We use superscripts on $\bm{r}$ to distinguish the 3D positions of atoms in different iterations. for example, at the $i$-th iteration, the update can be denoted as $ (y, \bm{r}^{(i)} ) = f(\bm{X}, \bm{E}, \bm{r}^{(i-1)} ; \bm{\theta})$. 
It is noteworthy that parameters $\bm{\theta}$ are shared across all iterations. Moreover, please note that the iterative update in \name involves only a few rounds, such as 1 or 2, instead of dozens or hundreds of steps in Molecular Dynamics.

\subsection{Model Training} \label{sec:training}

In DFT conformation optimization or Molecular Dynamics simulations, a conformation is optimized step-by-step, resulting in a trajectory from a raw conformation to the equilibrium conformation in Euclidean space. However, saving such a trajectory can be expensive, and publicly available datasets usually provide the equilibrium conformations only. Providing a trajectory would be beneficial as intermediate states can be used as data augmentation to guide the model's training. Inspired by this, we propose a novel training approach, which generates a pseudo trajectory first,  samples a conformation from it, and uses the sampled conformation as input to predict the equilibrium conformation. This approach allows us to better exploit the information in the molecular data, which we found can greatly improve the model's performance. Specifically, we assume that the trajectory from a raw conformation $\bm{r}^{\text{init}}$ to a target equilibrium conformation $\bm{r}^{\text{tgt}}$ is a linear process. We generate an intermediate conformation along this trajectory via noisy interpolation, i.e.,
\begin{equation}
\small
    \bm{r}^{(0)} = q \bm{r}^{\text{init}} + (1-q) \left(\bm{r}^{\text{tgt}} + \bm{c} \right),
\end{equation}
where scalar $q$ ranges from 0 to 1, the Gaussian noise $\bm{c} \in \mathbb{R}^{n \times 3}$ has a mean of 0 and standard deviation $\upsilon$ (a hyper-parameter). Taking $\bm{r}^{(0)}$ as input, \name learns to update towards the target equilibrium conformation $\bm{r}^{\text{tgt}}$. During inference, $q$ is set to $1.0$ by default. To address the distributional shift between training and inference (i.e., prefer $q=1.0$) and enhance the learning of an accurate mapping from the equilibrium conformation to the QC properties (i.e., prefer $q=0.0$), we use a mixture of Bernoulli distribution and Uniform distribution. Therefore we can flexibly assign higher sample probabilities to $q=1.0$ and $q=0.0$, as well as sampling from interpolations. The model takes $\bm{r}^{(0)}$ as input and generates $\bm{r}^{(R)}$ after $R$ iterations. Then, the model uses $\bm{r}^{(R)}$ as input and predicts the QC properties. L1 loss is applied to the QC property regression and the 3D coordinate prediction. \looseness=-1

\section{Experiment}

This section provides an empirical analysis of the performance of \name. First, the configuration of \name is outlined, followed by a thorough benchmarking using two widely-recognized public datasets: PCQM4MV2 \cite{DBLP:conf/nips/HuFRNDL21} and OC20 \cite{chanussot2021open}. These datasets facilitate the evaluation of \name's performance in small organic molecules and catalyst systems. Finally, an ablation study is conducted to examine the influence of various model components and training strategies on the overall performance.

\subsection{Model Configuration}

Similar to both Graphormer~\cite{ying2021transformers} and Transformer-M~\cite{luo2022one}, \name comprises 12 layers with an atom representation dimension of $d_x=768$ and a pair representation dimension of $d_p=256$. 
The hidden dimension of \ffn in the atom representation track is set to $768$, while that of the pair representation track is set to $256$. Additionally, the hidden dimension in the \opm is $d_o=32$, and the hidden dimension in the \tm is $d_t=32$ as well. The number of conformation optimization iterations $R$ is set to $1$, indicating that the model iterates twice in total (once for conformation optimization and once for quantum chemistry property prediction).
For the training strategy, we specified a standard deviation of $\upsilon = 0.2$ for random noise and employed a particular sampling method for $q$. Specifically, $q$ was set to $0.0$ with probability $0.8$, set to $1.0$ with probability $0.1$, and uniformly sampled from $[0.4, 0.6]$ with probability $0.1$. 
With this setting, the number of parameters of \name is about 52.4M. \looseness=-1

\subsection{PCQM4MV2}

\begin{table}[t]
  \small
  \centering
  \caption{Performance on PCQM4MV2.} \label{tab:pcq}
  %\addtolength{\tabcolsep}{-3pt}
    \begin{tabular}{l|c|c|c|c}
    \toprule
    Model & \# param. & \# layers & Valid MAE ($\downarrow$) & Leaderboard MAE ($\downarrow$)  \\
    \midrule
    MLP-Fingerprint~\cite{DBLP:conf/nips/HuFRNDL21} & 16.1M &- & 0.1735 & 0.1760	 \\
    GCN~\cite{kipf2016semi} & 2.0M &- & 0.1379 & 0.1398 \\
    GIN~\cite{xu2018powerful} & 3.8M &- & 0.1195 & 0.1218 \\
    \text{GINE}-$_\text{VN}$~\cite{brossard2020graph, gilmer2017neural,luo2022one} & 13.2M &- & 0.1167 & -\\
    \text{GCN}-$_\text{VN}$~\cite{kipf2016semi, gilmer2017neural} & 4.9M &- & 0.1153 & 0.1152	 \\
    \text{GIN}-$_\text{VN}$~\cite{xu2018powerful, gilmer2017neural} & 6.7M &- & 0.1083 & 0.1084 \\
    \text{DeeperGCN}-$_\text{VN}$~\cite{li2020deepergcn, luo2022one} & 25.5M & 12 & 0.1021 & - \\
    $\text{GraphGPS}_\text{SMALL}$~\cite{rampavsek2022recipe} & 6.2M & 5 & 0.0938 & -\\
    TokenGT~\cite{kim2022pure} & 48.5M & 12 & 0.0910 & 0.0919 \\
    $\text{GRPE}_\text{BASE}$~\cite{park2022grpe} & 46.2M & 12 & 0.0890 & -\\
    EGT~\cite{hussain2022global} & 89.3M & 24 & 0.0869 & 0.0872 \\
    $\text{GRPE}_\text{LARGE}$~\cite{hussain2022global} & 46.2M & 18 & 0.0867 & 0.0876 \\
    Graphormer~\cite{ying2021transformers, shi2022benchmarking} & 47.1M & 12 & 0.0864 & -\\
    $\text{GraphGPS}_\text{BASE}$~\cite{rampavsek2022recipe} & 19.4M & 10 & 0.0858 & - \\
    $\text{GraphGPS}_\text{DEEP}$~\cite{rampavsek2022recipe} & 13.8M & 16 & 0.0852 & 0.0862 \\
    GEM-2~\cite{liu2022gem} & 32.1M & 12 & 0.0793 &0.0806 \\
    GPS++~\cite{masters2022gps++} & 44.3M & 16 &  0.0778 & 0.0720 \footnotemark[1] \\
    \midrule
    \multirow{2}{*}{Transformer-M~\cite{luo2022one}} & 47.1M & 12 & 0.0787 & -\\
     & 69M & 18 & 0.0772 & 0.0782  \\
    \midrule
    \multirow{3}{*}{\name} & 27.7M & 6 & 0.0714  & -\\
    & 52.4M & 12 &  0.0696 & -\\
     & 77M &  18 & \textbf{0.0693} &  \textbf{0.0705} \\
    \bottomrule
    \end{tabular}
    \vspace{-6pt}
\end{table}

The PCQM4Mv2 dataset, derived from the OGB Large-Scale Challenge \cite{DBLP:conf/nips/HuFRNDL21}, is designed to facilitate the development and evaluation of machine learning models for predicting QC properties of molecules, specifically the target property known as the "HOMO-LUMO gap." This property represents the difference between the energies of the highest occupied molecular orbital (HOMO) and the lowest unoccupied molecular orbital (LUMO).
The dataset, consisting of approximately 4 million molecules represented by SMILES notations, offers HOMO-LUMO gap labels for the training and validation sets; however, the labels for the test set remain undisclosed. Furthermore, the training set encompasses the DFT equilibrium conformation, which is not included in the validation and test sets. The benchmark's goal is to utilize SMILES notation, without the DFT equilibrium conformation, to predict the HOMO-LUMO gap during the inference process.

\vspace{-6pt}
\paragraph{Setting}
Based on SMILES, we generated 8 initial conformations for each molecule by RDKit, at a per-molecule cost of about 0.01 seconds. During training, we randomly sampled 1 conformation as input $\bm{r}$ at each epoch, while during inference, we used the average HOMO-LUMO gap prediction based on 8 conformations. We used the AdamW optimizer with a learning rate of $2e\text{-}4$, a batch size of $1024$, $(\beta_1, \beta_2)$ set to $(0.9, 0.999)$, and gradient clipping set to $5.0$ during training, which lasted for 1.5 million steps, with 150K warmup steps. Additionally, an exponential moving average (EMA) with a decay rate of 0.999 was utilized. 
The training took approximately 5 days, utilizing 8 NVIDIA A100 GPUs. 
The inference on the 147k test-dev set took approximately 7 minutes, utilizing 8 NVIDIA V100 GPUs. 

We incorporate previous submissions to the PCQM4MV2 leaderboard as baselines. In addition to the default 12-layer model, we evaluate the performance of \name with two variants consisting of 6 and 18 layers, respectively, to investigate the impact of varying model parameter sizes.

\vspace{-6pt}
\paragraph{Result} 
The results are summarized in Table~\ref{tab:pcq}, and our observations are as follows: (1) All three variants of \name demonstrate substantial performance improvements over previous baselines. (2) The 6-layer \name, despite having considerably fewer model parameters, outperforms all prior baselines. (3) Increasing the layers from 6 to 12 results in a significant accuracy enhancement, surpassing all baselines by a considerable margin. (4) The 18-layer \name exhibits the highest performance, outperforming all baselines by a remarkable margin. These findings underscore the effectiveness of \name. (5) The performance of a single 18-layer \name model on the leaderboard (test-dev set) is noteworthy, particularly as it surpasses previous state-of-the-art methods without employing an ensemble or additional techniques. In contrast, the previous state-of-the-art GPS++ relied on a 112-model ensemble and included the validation set for training.

\footnotetext[1]{GPS++'s leaderboard submission consists of a 112-model ensemble and utilizes the validation data for training.}

\subsection{Open Catalyst 2020}

The Open Catalyst 2020 (OC20) dataset \cite{chanussot2021open} is specifically designed to promote the development of machine-learning models for catalyst discovery and optimization. OC20 encompasses three tasks: Structure to Energy and Force (S2EF), Initial Structure to Relaxed Structure (IS2RS), and Initial Structure to Relaxed Energy (IS2RE). In this paper, we focus on the IS2RE task, as it aligns well with the objectives of the proposed methodology. The goal of the IS2RE task is to predict the relaxed energy based on the initial conformation. It comprises approximately 460K training data points. While DFT equilibrium conformations are provided for training, they are not permitted for use during inference. Moreover, in contrast to the PCQM4MV2 dataset, the initial conformation is already supplied in the OC20 IS2RE task, eliminating the need to generate the initial input conformation by ourselves. 

\vspace{-6pt}
\paragraph{Setting}

We use the default 12-layer \name setting for OC20 experiments. The model configuration deviates slightly from the settings employed in PCQM4MV2. Firstly, since OC20 lacks graph information, graph-related features are excluded from the model. Secondly, due to the greater number of atoms present in OC20 compared to PCQM4MV2, the model capacity is marginally reduced for efficiency reasons. In particular, the pair representation dimension $d_p$ is set to $128$, while the hidden dimensions in the \opm and \tm are set to $d_o=16$ and $d_t=16$, respectively. 
Third, the periodic boundary condition needs to be considered; we adopt the solution proposed in \cite{shi2022benchmarking}, which pre-expands the neighbor cells and then applies a radius cutoff to reduce the number of atoms.
The AdamW optimizer was employed during the training process, which lasted for 1.5 million steps, including 150K warmup steps. The optimizer was configured with a learning rate of $2e\text{-}4$, a batch size of $64$, $(\beta_1, \beta_2)$ values of $(0.9, 0.999)$, and a gradient clipping parameter of $5.0$. 
The training process spanned approximately 7 days and made use of 16 NVIDIA A100 GPUs.

\vspace{-6pt}
\paragraph{Result} 

We present a performance comparison of various models on the OC20 IS2RE validation and test set, as illustrated in Table~\ref{tab:oc20} and Table~\ref{tab:oc20_test}. The two tables display the Mean Absolute Error (MAE) for energy in electron volts (eV) and the percentage of Energies Within a Threshold (EwT) for each model. As evident from the tables, our proposed \name significantly outperforms all previous baselines in terms of both MAE and EwT, demonstrating its exceptional performance. Notably, our method attains the lowest MAEs across all categories, including In-Domain (ID), Out-of-Domain Adsorption (OOD Ads.), Out-of-Domain Catalysis (OOD Cat.), Out-of-Domain Both (OOD Both), and Average (AVG.). Furthermore, in terms of EwT, \name consistently achieves the highest values in all categories. These findings underscore the robustness of our method in handling both in-domain and out-of-domain data.
In conclusion, the results emphasize the efficacy of our approach in capturing intricate interactions in material systems and its potential for extensive applicability in various computational material science tasks.

\begin{table}[t]
  \small
  \caption{Performance on OC20 IS2RE validation set. NN refers to "Noisy Nodes"\cite{godwin2022simple}.} \label{tab:oc20}
  \addtolength{\tabcolsep}{-3.3pt}
    \begin{tabular}{l|ccccc|ccccc}
    \toprule
    & \multicolumn{5}{|c|}{Energy MAE (eV) $\downarrow$} & \multicolumn{4}{|c}{ EwT (\%) $\uparrow$}  \\
    Model & {\scriptsize ID} & {\scriptsize OOD Ads. }& {\scriptsize OOD Cat.} & {\scriptsize OOD Both} & {\scriptsize AVG.} & {\scriptsize ID} & {\scriptsize OOD Ads.} & {\scriptsize OOD Cat.} &  {\scriptsize OOD Both} & {\scriptsize AVG.}  \\
    \midrule
    SchNet~\cite{schutt2017schnet} & 0.6465 & 0.7074 & 0.6475 & 0.6626 & 0.6660 & 2.96 & 2.22 & 3.03 & 2.38 & 2.65\\
    DimeNet++~\cite{gasteiger_dimenetpp_2020} & 0.5636 & 0.7127 & 0.5612 & 0.6492 & 0.6217 & 4.25 & 2.48 & 4.40 & 2.56 & 3.42\\
    GemNet-T~\cite{GasteigerBG21} & 0.5561 & 0.7342 & 0.5659 & 0.6964 & 0.6382 & 4.51 & 2.24 & 4.37 & 2.38 & 3.38\\
    SphereNet~\cite{0059WLLZOJ22} & 0.5632 & 0.6682 & 0.5590 & 0.6190 & 0.6024 & 4.56 & 2.70 & 4.59 & 2.70 & 3.64\\
    \midrule
    Graphormer-3D~\cite{shi2022benchmarking} & 0.4329 & 0.5850 & 0.4441 & 0.5299 & 0.4980 & - & - & - & - & -	 \\
    GNS~\cite{godwin2022simple} & 0.54 & 0.65 & 0.55 & 0.59 & 0.5825 & - & - & - & - & -  \\
    GNS+NN~\cite{godwin2022simple} & 0.47 & 0.51 & 0.48 & 0.46 & 0.4800 & - & - & - & - & - \\
    EquiFormer~\cite{liao2022equiformer} & 0.4222 & 0.5420 & 0.4231 & 0.4754 & 0.4657 & 7.23 & 3.77 & 7.13 & 4.10 & 5.56  \\
    EquiFormer+NN~\cite{liao2022equiformer} & 0.4156 & 0.4976 & 0.4165 & 0.4344 & 0.4410 & 7.47 & 4.64 & 7.19 & 4.84 & 6.04 \\
    DRFormer~\cite{wang2023dr} & 0.4187& 0.4863 & 0.4321 & 0.4332 & 0.4425 & 8.39 & 5.42 & 8.12 & 5.44 & 6.84 \\
    \midrule
    \name & \textbf{0.3795} & \textbf{0.4526} & \textbf{0.4011} & \textbf{0.4021} & \textbf{0.4088} & \textbf{11.15} & \textbf{6.71} & \textbf{9.90} & \textbf{6.68} & \textbf{8.61} \\
    \bottomrule
    \end{tabular}
    \vspace{-6pt}
\end{table}

\begin{table}[t]
  \small
  \caption{Performance on OC20 IS2RE test set. NN refers to "Noisy Nodes"\cite{godwin2022simple}.} \label{tab:oc20_test}
  \addtolength{\tabcolsep}{-3.3pt}
    \begin{tabular}{l|ccccc|ccccc}
    \toprule
    & \multicolumn{5}{|c|}{Energy MAE (eV) $\downarrow$} & \multicolumn{4}{|c}{ EwT (\%) $\uparrow$}  \\
    Model & {\scriptsize ID} & {\scriptsize OOD Ads. }& {\scriptsize OOD Cat.} & {\scriptsize OOD Both} & {\scriptsize AVG.} & {\scriptsize ID} & {\scriptsize OOD Ads.} & {\scriptsize OOD Cat.} &  {\scriptsize OOD Both} & {\scriptsize AVG.}  \\
    \midrule
    SchNet~\cite{schutt2017schnet} & 0.639 & 0.734 & 0.662 &  0.704 & 0.6848  & 2.96 & 2.33 & 2.94 & 2.21 & 2.61 \\
    DimeNet++~\cite{gasteiger_dimenetpp_2020} & 0.562 & 0.725 & 0.576 &  0.661 & 0.631 & 4.25 & 2.07 & 4.1 & 2.41 & 3.21 \\
    SphereNet~\cite{0059WLLZOJ22} & 0.563 & 0.703 & 0.571 & 0.638 & 0.6188 & 4.47 & 2.29 & 4.09 & 2.41 & 3.32 \\
    \midrule
    Graphormer-3D~\cite{shi2022benchmarking} & 0.3976 & 0.5719 & 0.4166 & 0.5029 & 0.4722 & 8.97 & 3.45 & 8.18 & 3.79 & 6.1 \\
    GNS+NN~\cite{godwin2022simple} & 0.4219 & 0.5678 & 0.4366 & 0.4651 & 0.4728 & 9.12 & 4.25 & 8.01 & 4.64 & 6.5 \\
    EquiFormer~\cite{liao2022equiformer} & 0.5037 & 0.6881 & 0.5213 & 0.6301 & 0.5858 & 5.14 & 2.41 & 4.67 & 2.69 & 3.73 \\
    EquiFormer+NN~\cite{liao2022equiformer} & 0.4171 & 0.5479 & 0.4248 & 0.4741 & 0.4660 & 7.71 & 3.70 & 7.15 & 4.07 & 5.66 \\
    DRFormer~\cite{wang2023dr} & 0.3865 & 0.5435 & 0.4060 & 0.4677 & 0.4509 & 9.18 & 4.01 & 8.39 & 4.33 & 6.48 \\
    \midrule
    \name & \textbf{0.3745} & \textbf{0.4760} & \textbf{0.3980} & \textbf{0.4086} & \textbf{0.4143} & \textbf{11.29} & \textbf{6.05} & \textbf{9.53} & \textbf{6.06} & \textbf{8.23} \\
    \bottomrule
    \end{tabular}
    \vspace{-6pt}
\end{table}

\vspace{-6pt}
\section{Conclusion}
\vspace{-6pt}

In this paper, we presented \name, a novel method for predicting quantum chemical properties using deep learning. Unlike previous methods that directly predict QC properties from 1D/2D data, our approach first generates raw 3D conformations from 1D/2D data using cost-effective methods and then iteratively updates these conformations towards DFT equilibrium conformations using neural networks. To facilitate effective learning of the conformation optimization process, we developed a new two-track Transformer backbone and a novel training strategy. We evaluated \name on multiple public benchmark datasets, including PCQM4MV2 and Open Catalyst 2020, where it outperformed all previous state-of-the-art methods by a significant margin. Ablation studies further confirmed the effectiveness of each component in our proposed method.

In conclusion, our work demonstrates the potential of leveraging deep learning to accelerate quantum chemical property prediction by refining raw 3D conformations towards DFT equilibrium conformations. This approach holds promise for enabling more efficient high-throughput screening and design of novel materials and molecules in the future.

\vspace{-6pt}
\section*{Limitations}
\vspace{-6pt}

The proposed method is primarily developed for quantum chemical property prediction, wherein equilibrium conformations optimized through electronic structure methods can be obtained during dataset construction. Nevertheless, the method may not perform optimally for molecular properties that cannot be calculated using electronic structure methods, such as toxicity.

\bibliographystyle{plain}
\bibliography{ref}

\begin{thebibliography}{10}

\bibitem{brossard2020graph}
R{\'e}my Brossard, Oriel Frigo, and David Dehaene.
\newblock Graph convolutions that can finally model local structure.
\newblock {\em arXiv preprint arXiv:2011.15069}, 2020.

\bibitem{chanussot2021open}
Lowik Chanussot, Abhishek Das, Siddharth Goyal, Thibaut Lavril, Muhammed
  Shuaibi, Morgane Riviere, Kevin Tran, Javier Heras-Domingo, Caleb Ho, Weihua
  Hu, et~al.
\newblock Open catalyst 2020 (oc20) dataset and community challenges.
\newblock {\em Acs Catalysis}, 11(10):6059--6072, 2021.

\bibitem{GasteigerBG21}
Johannes Gasteiger, Florian Becker, and Stephan G{\"{u}}nnemann.
\newblock Gemnet: Universal directional graph neural networks for molecules.
\newblock In Marc'Aurelio Ranzato, Alina Beygelzimer, Yann~N. Dauphin, Percy
  Liang, and Jennifer~Wortman Vaughan, editors, {\em Advances in Neural
  Information Processing Systems 34: Annual Conference on Neural Information
  Processing Systems 2021, NeurIPS 2021, December 6-14, 2021, virtual}, pages
  6790--6802, 2021.

\bibitem{gasteiger_dimenetpp_2020}
Johannes Gasteiger, Shankari Giri, Johannes~T. Margraf, and Stephan
  G{\"u}nnemann.
\newblock Fast and uncertainty-aware directional message passing for
  non-equilibrium molecules.
\newblock In {\em Machine Learning for Molecules Workshop, NeurIPS}, 2020.

\bibitem{gilmer2017neural}
Justin Gilmer, Samuel~S Schoenholz, Patrick~F Riley, Oriol Vinyals, and
  George~E Dahl.
\newblock Neural message passing for quantum chemistry.
\newblock In {\em International conference on machine learning}, pages
  1263--1272. PMLR, 2017.

\bibitem{godwin2022simple}
Jonathan Godwin, Michael Schaarschmidt, Alexander~L Gaunt, Alvaro
  Sanchez-Gonzalez, Yulia Rubanova, Petar Veli{\v{c}}kovi{\'c}, James
  Kirkpatrick, and Peter Battaglia.
\newblock Simple {GNN} regularisation for 3d molecular property prediction and
  beyond.
\newblock In {\em International Conference on Learning Representations}, 2022.

\bibitem{gomez2018automatic}
Rafael G{\'o}mez-Bombarelli, Jennifer~N Wei, David Duvenaud, Jos{\'e}~Miguel
  Hern{\'a}ndez-Lobato, Benjam{\'\i}n S{\'a}nchez-Lengeling, Dennis Sheberla,
  Jorge Aguilera-Iparraguirre, Timothy~D Hirzel, Ryan~P Adams, and Al{\'a}n
  Aspuru-Guzik.
\newblock Automatic chemical design using a data-driven continuous
  representation of molecules.
\newblock {\em ACS central science}, 4(2):268--276, 2018.

\bibitem{guan2022energyinspired}
Jiaqi Guan, Wesley~Wei Qian, qiang liu, Wei-Ying Ma, Jianzhu Ma, and Jian Peng.
\newblock Energy-inspired molecular conformation optimization.
\newblock In {\em International Conference on Learning Representations}, 2022.

\bibitem{DBLP:conf/nips/HuFRNDL21}
Weihua Hu, Matthias Fey, Hongyu Ren, Maho Nakata, Yuxiao Dong, and Jure
  Leskovec.
\newblock {OGB-LSC:} {A} large-scale challenge for machine learning on graphs.
\newblock In Joaquin Vanschoren and Sai{-}Kit Yeung, editors, {\em Proceedings
  of the Neural Information Processing Systems Track on Datasets and Benchmarks
  1, NeurIPS Datasets and Benchmarks 2021, December 2021, virtual}, 2021.

\bibitem{hussain2022global}
Md~Shamim Hussain, Mohammed~J Zaki, and Dharmashankar Subramanian.
\newblock Global self-attention as a replacement for graph convolution.
\newblock In {\em Proceedings of the 28th ACM SIGKDD Conference on Knowledge
  Discovery and Data Mining}, pages 655--665, 2022.

\bibitem{hutchinson2021lietransformer}
Michael~J Hutchinson, Charline Le~Lan, Sheheryar Zaidi, Emilien Dupont,
  Yee~Whye Teh, and Hyunjik Kim.
\newblock Lietransformer: Equivariant self-attention for lie groups.
\newblock In {\em International Conference on Machine Learning}, pages
  4533--4543. PMLR, 2021.

\bibitem{jain2013commentary}
Anubhav Jain, Shyue~Ping Ong, Geoffroy Hautier, Wei Chen, William~Davidson
  Richards, Stephen Dacek, Shreyas Cholia, Dan Gunter, David Skinner, Gerbrand
  Ceder, et~al.
\newblock Commentary: The materials project: A materials genome approach to
  accelerating materials innovation.
\newblock {\em APL materials}, 1(1):011002, 2013.

\bibitem{jones2015density}
Robert~O Jones.
\newblock Density functional theory: Its origins, rise to prominence, and
  future.
\newblock {\em Reviews of modern physics}, 87(3):897, 2015.

\bibitem{jumper_highly_2021}
John Jumper, Richard Evans, Alexander Pritzel, Tim Green, Michael Figurnov,
  Olaf Ronneberger, Kathryn Tunyasuvunakool, Russ Bates, Augustin
  {\v{Z}}{\'\i}dek, Anna Potapenko, et~al.
\newblock Highly accurate protein structure prediction with alphafold.
\newblock {\em Nature}, 596(7873):583--589, 2021.

\bibitem{kim2022pure}
Jinwoo Kim, Tien~Dat Nguyen, Seonwoo Min, Sungjun Cho, Moontae Lee, Honglak
  Lee, and Seunghoon Hong.
\newblock Pure transformers are powerful graph learners.
\newblock {\em arXiv preprint arXiv:2207.02505}, 2022.

\bibitem{kipf2016semi}
Thomas~N Kipf and Max Welling.
\newblock Semi-supervised classification with graph convolutional networks.
\newblock {\em arXiv preprint arXiv:1609.02907}, 2016.

\bibitem{klicpera2003directional}
Johannes Klicpera, Janek Gro{\ss}, and Stephan G{\"u}nnemann.
\newblock Directional message passing for molecular graphs 2020.
\newblock {\em arXiv preprint arXiv:2003.03123}, 2003.

\bibitem{landrum2016rdkit}
Greg Landrum et~al.
\newblock Rdkit: Open-source cheminformatics software.
\newblock 2016.

\bibitem{li2020deepergcn}
Guohao Li, Chenxin Xiong, Ali Thabet, and Bernard Ghanem.
\newblock Deepergcn: All you need to train deeper gcns.
\newblock {\em arXiv preprint arXiv:2006.07739}, 2020.

\bibitem{liao2022equiformer}
Yi-Lun Liao and Tess Smidt.
\newblock Equiformer: Equivariant graph attention transformer for 3d atomistic
  graphs.
\newblock {\em arXiv preprint arXiv:2206.11990}, 2022.

\bibitem{liu2022gem}
Lihang Liu, Donglong He, Xiaomin Fang, Shanzhuo Zhang, Fan Wang, Jingzhou He,
  and Hua Wu.
\newblock Gem-2: Next generation molecular property prediction network with
  many-body and full-range interaction modeling.
\newblock {\em arXiv preprint arXiv:2208.05863}, 2022.

\bibitem{liu2021pre}
Shengchao Liu, Hanchen Wang, Weiyang Liu, Joan Lasenby, Hongyu Guo, and Jian
  Tang.
\newblock Pre-training molecular graph representation with 3d geometry.
\newblock {\em arXiv preprint arXiv:2110.07728}, 2021.

\bibitem{0059WLLZOJ22}
Yi~Liu, Limei Wang, Meng Liu, Yuchao Lin, Xuan Zhang, Bora Oztekin, and
  Shuiwang Ji.
\newblock Spherical message passing for 3d molecular graphs.
\newblock In {\em The Tenth International Conference on Learning
  Representations, {ICLR} 2022, Virtual Event, April 25-29, 2022}.
  OpenReview.net, 2022.

\bibitem{luo2022one}
Shengjie Luo, Tianlang Chen, Yixian Xu, Shuxin Zheng, Tie-Yan Liu, Liwei Wang,
  and Di~He.
\newblock One transformer can understand both 2d \& 3d molecular data.
\newblock {\em arXiv preprint arXiv:2210.01765}, 2022.

\bibitem{luo2022your}
Shengjie Luo, Shanda Li, Shuxin Zheng, Tie-Yan Liu, Liwei Wang, and Di~He.
\newblock Your transformer may not be as powerful as you expect.
\newblock {\em arXiv preprint arXiv:2205.13401}, 2022.

\bibitem{masters2022gps++}
Dominic Masters, Josef Dean, Kerstin Klaser, Zhiyi Li, Sam Maddrell-Mander,
  Adam Sanders, Hatem Helal, Deniz Beker, Ladislav Ramp{\'a}{\v{s}}ek, and
  Dominique Beaini.
\newblock Gps++: An optimised hybrid mpnn/transformer for molecular property
  prediction.
\newblock {\em arXiv preprint arXiv:2212.02229}, 2022.

\bibitem{park2022grpe}
Wonpyo Park, Woong-Gi Chang, Donggeon Lee, Juntae Kim, et~al.
\newblock Grpe: Relative positional encoding for graph transformer.
\newblock In {\em ICLR2022 Machine Learning for Drug Discovery}, 2022.

\bibitem{rampavsek2022recipe}
Ladislav Ramp{\'a}{\v{s}}ek, Mikhail Galkin, Vijay~Prakash Dwivedi, Anh~Tuan
  Luu, Guy Wolf, and Dominique Beaini.
\newblock Recipe for a general, powerful, scalable graph transformer.
\newblock {\em arXiv preprint arXiv:2205.12454}, 2022.

\bibitem{ross2022large}
Jerret Ross, Brian Belgodere, Vijil Chenthamarakshan, Inkit Padhi, Youssef
  Mroueh, and Payel Das.
\newblock Large-scale chemical language representations capture molecular
  structure and properties.
\newblock {\em Nature Machine Intelligence}, 4(12):1256--1264, 2022.

\bibitem{schutt2017schnet}
Kristof Sch{\"u}tt, Pieter-Jan Kindermans, Huziel~Enoc Sauceda~Felix, Stefan
  Chmiela, Alexandre Tkatchenko, and Klaus-Robert M{\"u}ller.
\newblock Schnet: A continuous-filter convolutional neural network for modeling
  quantum interactions.
\newblock {\em Advances in neural information processing systems}, 30, 2017.

\bibitem{shi2022benchmarking}
Yu~Shi, Shuxin Zheng, Guolin Ke, Yifei Shen, Jiacheng You, Jiyan He, Shengjie
  Luo, Chang Liu, Di~He, and Tie-Yan Liu.
\newblock Benchmarking graphormer on large-scale molecular modeling datasets,
  2022.

\bibitem{stark20223d}
Hannes St{\"a}rk, Dominique Beaini, Gabriele Corso, Prudencio Tossou, Christian
  Dallago, Stephan G{\"u}nnemann, and Pietro Li{\`o}.
\newblock 3d infomax improves gnns for molecular property prediction.
\newblock In {\em International Conference on Machine Learning}, pages
  20479--20502. PMLR, 2022.

\bibitem{tholke2022equivariant}
Philipp Th{\"o}lke and Gianni De~Fabritiis.
\newblock Equivariant transformers for neural network based molecular
  potentials.
\newblock In {\em International Conference on Learning Representations}, 2022.

\bibitem{wang2023dr}
Bowen Wang, Chen Liang, Jiaze Wang, Furui Liu, Shaogang Hao, Dong Li, Jianye
  Hao, Guangyong Chen, Xiaolong Zou, and Pheng-Ann Heng.
\newblock Dr-label: Improving gnn models for catalysis systems by label
  deconstruction and reconstruction.
\newblock {\em arXiv preprint arXiv:2303.02875}, 2023.

\bibitem{wang2019smiles}
Sheng Wang, Yuzhi Guo, Yuhong Wang, Hongmao Sun, and Junzhou Huang.
\newblock Smiles-bert: large scale unsupervised pre-training for molecular
  property prediction.
\newblock In {\em Proceedings of the 10th ACM international conference on
  bioinformatics, computational biology and health informatics}, pages
  429--436, 2019.

\bibitem{xu2018powerful}
Keyulu Xu, Weihua Hu, Jure Leskovec, and Stefanie Jegelka.
\newblock How powerful are graph neural networks?
\newblock {\em arXiv preprint arXiv:1810.00826}, 2018.

\bibitem{ying2021transformers}
Chengxuan Ying, Tianle Cai, Shengjie Luo, Shuxin Zheng, Guolin Ke, Di~He,
  Yanming Shen, and Tie-Yan Liu.
\newblock Do transformers really perform badly for graph representation?
\newblock {\em Advances in Neural Information Processing Systems},
  34:28877--28888, 2021.

\bibitem{zhang2018deep}
Linfeng Zhang, Jiequn Han, Han Wang, Roberto Car, and Weinan E.
\newblock Deep potential molecular dynamics: a scalable model with the accuracy
  of quantum mechanics.
\newblock {\em Physical review letters}, 120(14):143001, 2018.

\bibitem{zhou2022uni}
Gengmo Zhou, Zhifeng Gao, Qiankun Ding, Hang Zheng, Hongteng Xu, Zhewei Wei,
  Linfeng Zhang, and Guolin Ke.
\newblock Uni-mol: A universal 3d molecular representation learning framework.
\newblock In {\em The Eleventh International Conference on Learning
  Representations}, 2023.

\end{thebibliography}

\appendix

\end{document}